\documentclass[journal=jpclcd,manuscript=article]{achemso}

\usepackage[version=3]{mhchem} 
\usepackage{braket}

\author{Albert Liu}
\affiliation{Department of Physics, University of Michigan, Ann Arbor, Michigan 48109, USA}

\author{Diogo B. Almeida}
\affiliation{Department of Physics, University of Michigan, Ann Arbor, Michigan 48109, USA}

\author{Wan-Ki Bae}
\affiliation{SKKU Advanced Institute of Nano Technology, Sungkyunkwan University, Gyeonggi, Republic of Korea}

\author{Lazaro A. Padilha}
\affiliation{Instituto de Fisica ``Gleb Wataghin", Universidade de Campinas, 13083-970 Campinas, Sao Paulo, Brazil}

\author{Steven T. Cundiff}
\email{cundiff@umich.edu}
\affiliation{Department of Physics, University of Michigan, Ann Arbor, Michigan 48109, USA}

\title{Simultaneous Existence of Confined and Delocalized Vibrational Modes in Colloidal Quantum Dots}

\begin{document}

\begin{abstract}
    Coupling to phonon modes is a primary mechanism of excitonic dephasing and energy loss in semiconductors. However, low-energy phonons in colloidal quantum dots and their coupling to excitons are poorly understood, since their experimental signatures are weak and usually obscured by unavoidable inhomogeneous broadening of colloidal dot ensembles. We use multi-dimensional coherent spectroscopy at cryogenic temperatures to extract the homogeneous nonlinear optical response of excitons in a CdSe/CdZnS core/shell colloidal quantum dot ensemble. Comparison to simulation provides evidence that the observed lineshapes arise from the co-existence of confined and delocalized vibrational modes, both of which couple strongly to excitons in CdSe/CdZnS colloidal quantum dots.
\end{abstract}

Semiconductor quantum dots, structures that confine electronic excitations in three dimensions, are a maturing technology that have potential applications in numerous areas of optoelectronics. These include areas such as single-photon emitters \cite{Yuan2002,Senellart2017}, quantum computing \cite{Loss1998,Kloeffel2013,Gazzano2016}, and photovoltaics \cite{Carey2015,Zhao2017}. 

Though development of practical quantum dot devices is progressing rapidly, most areas of quantum dot optoelectronics suffer from the common issue of vibrational coupling to excitons. Interactions between electronic excitations and lattice vibrations facilitate energy loss through non-radiative decay processes and play a vital role in coherent control protocols \cite{Quilter2015,Kaldewey2017,Reindl2017}. Acoustic vibrational modes are of particular importance, since their low energies mediate coupling between the fine-structure of an exciton manifold \cite{Poem2010,Masia2012,Luker2017-2} and comprise the primary dephasing mechanism of interband coherences \cite{Muljarov2004,Accanto2012,Jakubczyk2016} in quantum dots. Though usually a continuum of modes, acoustic vibrations can assume discrete modes \cite{Takagahara1993,Kuok2003} due to size-confinement in colloidal quantum dots \cite{Chilla2008,Oron2009,Huxter2010,Werschler2016}. However, the two pictures of acoustic vibrations as a bulk-like phonon continuum or discrete spherical harmonics are seldom considered simultaneously.

Vibrational coupling to excitons in quantum dots has been studied extensively, but most spectroscopic studies have utilized linear techniques such as absorption and photoluminescence. Linear spectrocopies encounter two main obstacles. First, ensembles of all types of quantum dots exhibit inhomogeneous broadening (due to dot size dispersion) of their absorption and emission profiles. Linear techniques only provide the inhomogeneous lineshape of a quantum dot ensemble that reflects its size distribution, and are largely insensitive to its microscopic dynamics. Second, single quantum dot spectroscopy studies that circumvent inhomogeneous broadening have inherent limitations such as time-resolution and dot-to-dot structural variations. These limitations have restricted the focus of most experimental studies to properties of discrete vibrational modes, since signatures of continuum mode coupling are generally weak. Studies that do observe continuum modes are heavily influenced by spectral diffusion at timescales shorter than the integration time \cite{Empedocles1999}, and vary greatly between dots \cite{Fernee2008}.

A technique capable of extracting the homogeneous response of an inhomogeneously broadened ensemble is multi-dimensional coherent spectroscopy (MDCS) \cite{Cundiff2013}, which correlates absorption, intraband (Raman) coherence \cite{Ferrio1998}, and emission spectra. MDCS has recently been applied to a variety of quantum dot systems, including interfacial \cite{Kasprzak2010,Martin2018}, self-assembled \cite{Langbein2013,Suzuki2016}, and colloidal \cite{Harel2012,Karki2014,Gellen2017,Park2017,Seiler2018,Liu2018} dots, to reveal physics normally obscured by inhomogeneous broadening and/or single-dot experiment limitations. MDCS provides material properties that are both ensemble-averaged and size-resolved for all resonance frequencies within the excitation bandwidth.

In this Letter, we use MDCS to study an ensemble of core/shell colloidal quantum dots (CQDs) at cryogenic temperatures. The third-order response that we measure via MDCS enhances vibrational lineshapes due to phonon coupling, allowing for a sensitive, direct characterization of acoustic phonon coupling in the material. We not only characterize coupling to discrete acoustic modes related to the dot geometry \cite{Kuok2003}, but also observe strong features characteristic of coupling to a continuum harmonic bath. We attribute this bath to continuum acoustic modes of the quantum dot lattice \cite{Sirenko1998,Murray2004}, and argue for the coexistence of both a bulk-like continuum and discrete torsional vibrations. By comparison to simulation, we characterize the continuum mode spectral density and coupling mechanism.

\begin{figure}
    \centering
    \includegraphics[width=0.5\textwidth]{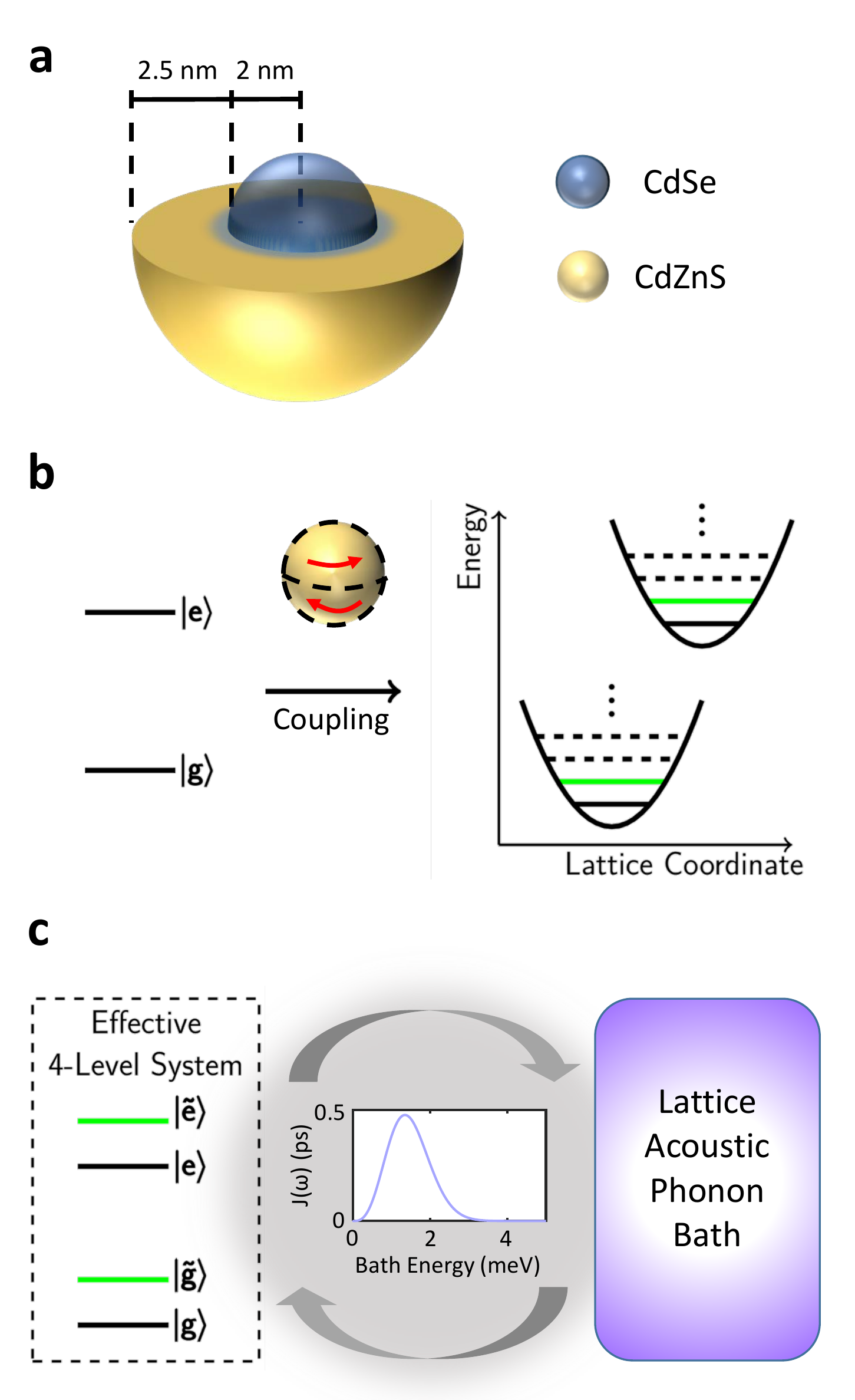}
    \caption{(a) Schematic of the CdSe/CdZnS core/shell colloidal quantum dots. (b) Dressing of the ground and excited exciton states by the $(\ell,n) = (2,0)$ torsional mode results in ladders of states separated by the (2,0) mode energy. (c) The simplified 4-level system formed from the lowest two states of the ground and excited state ladders then couple to a harmonic bath of acoustic phonon continuum states. The coupling strength is characterized by the spectral density function $J(\omega)$, and the $J(\omega)$ used in the simulations below is plotted inset.}
    \label{Fig1}
\end{figure}

The sample is CQDs composed of a 2 nm mean radius CdSe core and 2.5 nm mean thickness CdZnS shell (shown in Fig. \ref{Fig1}a), whose synthesis is detailed elsewhere \cite{Lim2014}. To study their properties at cryogenic temperatures the CQDs are dispersed in heptamethylnonane, which forms a transparent glass at temperatures below 100 K and is liquid up to room temperature. The colloidal suspension is diluted to an optical density of 0.3 at the room-temperature 1S exciton absorption peak.

The finite size of the CQD geometry introduces discrete vibrations \cite{Takagahara1993,Kuok2003}. These vibrations may be separated into the two classes of acoustic and torsional modes, which have longitudinal and transverse character respectively. However, CQDs embedded in a matrix may display an acoustic continuum characteristic of bulk materials \cite{Murray2004} if the sound velocity mismatch between the CQD and surrounding matrix is small \cite{Sirenko1998}, since boundary conditions at the CQD surface may be modified such that vibrations of the combined matrix and embedded sphere must be jointly considered.

MDCS is ideal to investigate acoustic phonon coupling in CQDs for two main reasons. First, the ensemble-averaged homogeneous response may be retrieved in the presence of inhomogeneity as a function of resonance energy (corresponding to radius in CQDs). Second, vibrational lineshapes are enhanced by nonlinear spectroscopies such as MDCS. 

A MDCS spectrum is generated by Fourier transforming a four-wave mixing signal along a combination of the two delays $\tau$ and $T$ between three excitation pulses and the evolution time $t$ after the last pulse. We acquire single-quantum spectra \cite{Hamm2011,Liu2018} by Fourier transforming along $\tau$ and $t$ \cite{Bristow2009}, which correlate absorption and emission spectra of the sample. Rephasing single-quantum spectra circumvent inhomogeneity by rephasing the evolution of excited coherences (due to a phase-conjugated first pulse \cite{Hamm2011}), as reflected in the negative absorption energies $\hbar\omega_\tau$. Cross-diagonal slices (taken perpendicular to the diagonal line $|\hbar\omega_\tau| = |\hbar\omega_t|$) provide the homogeneous third-order response.

To perform MDCS, we use a Multi-Dimensional Optical Nonlinear Spectrometer (MONSTR) \cite{Bristow2009}. 90 fs pulses at a 250 kHz repetition rate are split into four identical copies that are independently delayed in time and arranged in the box geometry. An excitation intensity of 4 W/cm$^2$ generates a predominately third-order response as verified by the power-dependence of the heterodyned signal.  All pulses are co-linearly polarized and centered at wavelength 605 nm.

\begin{figure*}
    \centering
    \includegraphics[width=1\textwidth]{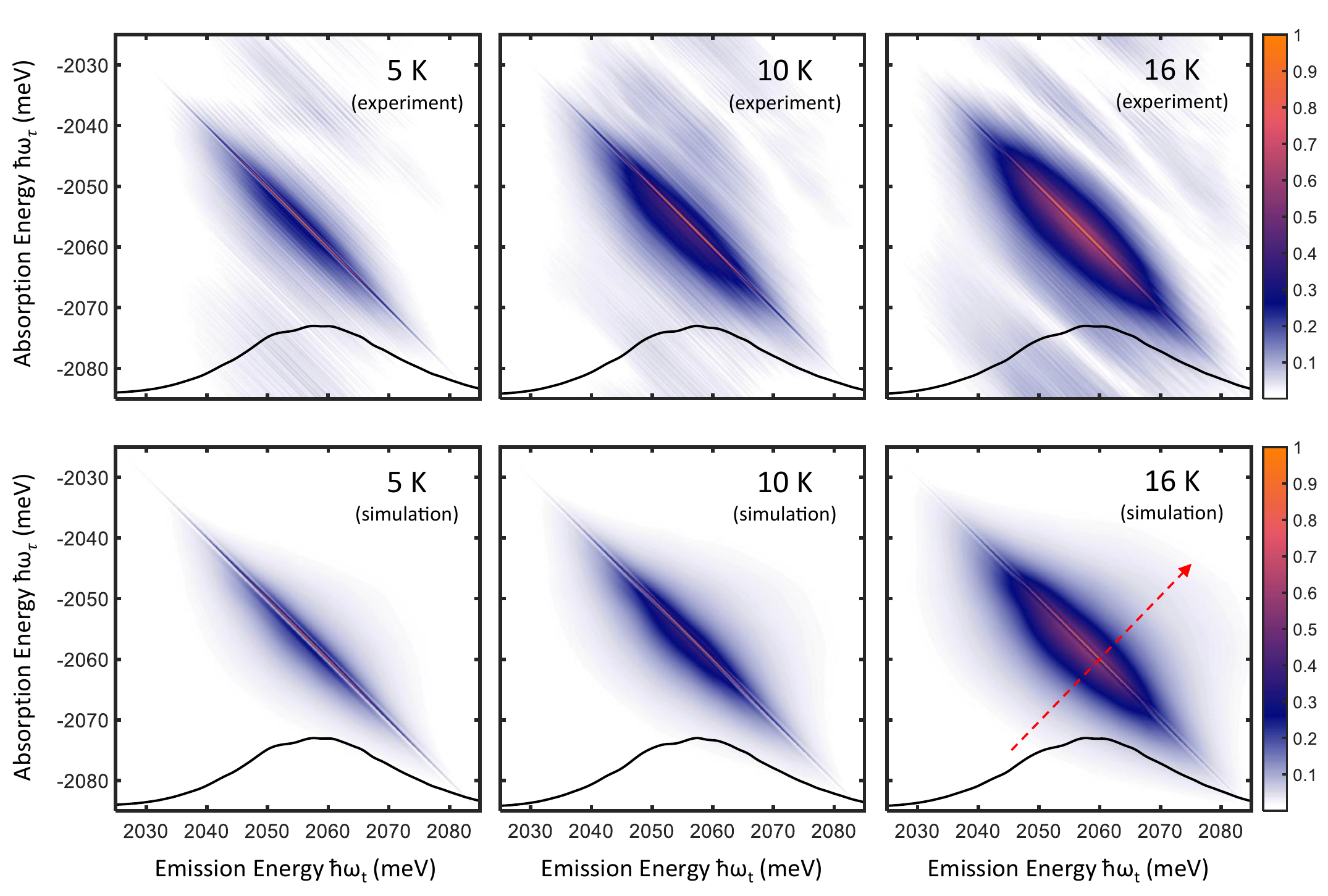}
    \caption{Experimental (top row) and simulated (bottom row) single-quantum spectra at temperatures 5, 10, and 16 K. All spectra are taken at waiting time $T = 1$ ps. The solid black curves represent the spectra of the excitation and local oscillator pulses for each temperature. A red dashed arrow in the bottom right panel indicates the location of the slices shown in Fig. \ref{Fig3}.}
    \label{Fig2}
\end{figure*}

\begin{figure}[H]
    \centering
    \includegraphics[width=0.5\textwidth]{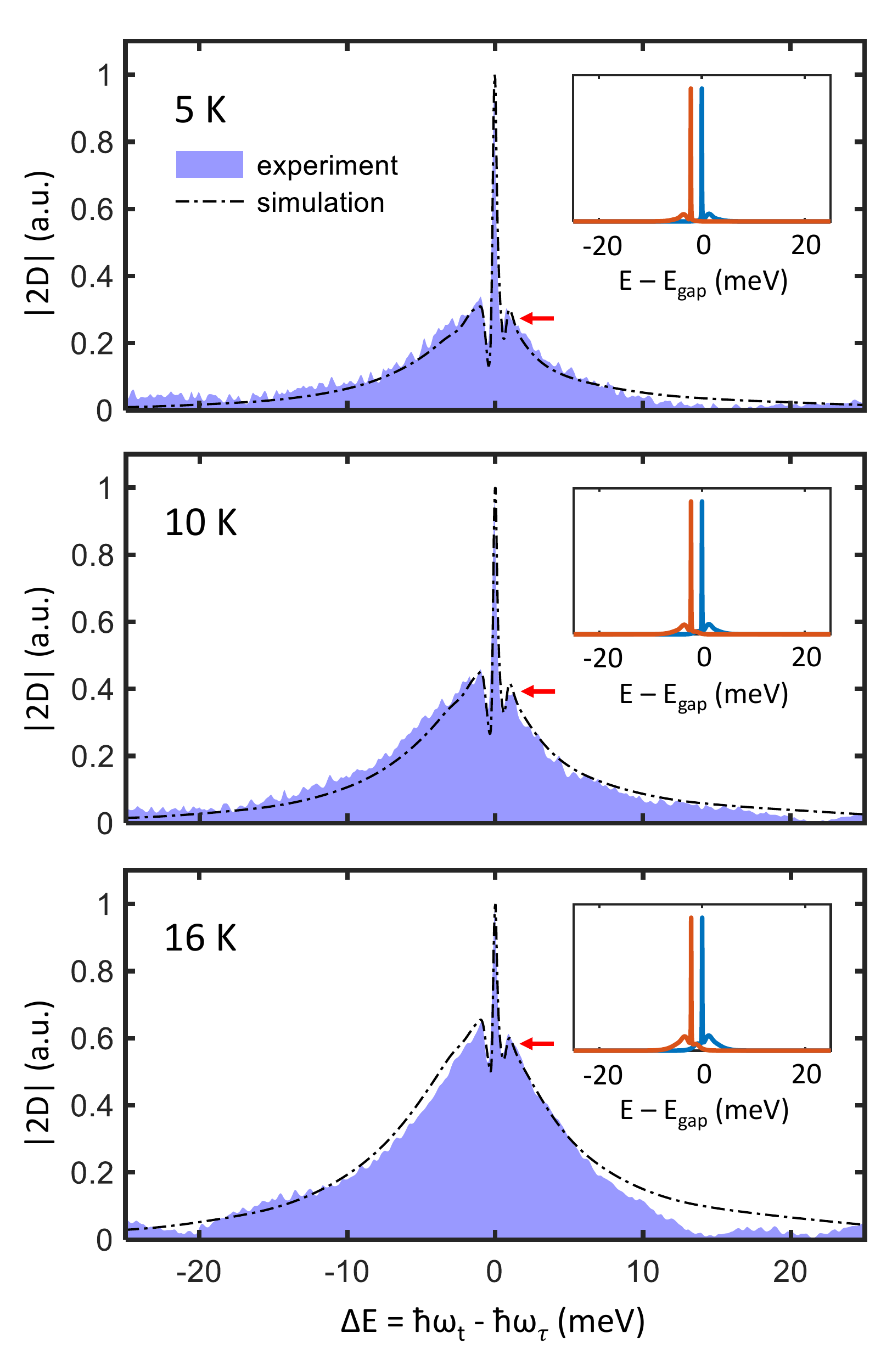}
    \caption{Cross-diagonal slices of the experimental (solid purple curve) and simulated (black line) single-quantum spectra at temperatures 5, 10, and 16 K. The slice locations are indicated by the red dashed arrow in Fig. \ref{Fig2}. The corresponding simulated absorption (blue) and fluorescence (orange) lineshapes are plotted inset as a function of detuning from the thermally averaged ground state electronic energy gap $E_{gap}$.}
    \label{Fig3}
\end{figure}

Single-quantum spectra are shown in the top row of Fig. \ref{Fig2} and have three main features. First, a narrow zero-phonon line is present along the diagonal, which corresponds to absorption and emission at the same energy. Second, a prominent broad pedestal around the zero-phonon line \cite{Palinginis2001,Palinginis2003}, which has the characteristic lineshape of localized excitons coupling to an acoustic phonon continuum bath \cite{Krummheuer2002,Besombes2001}, grows with increasing temperature. At these low temperatures, the pedestal is asymmetric due to higher probability of emission than absorption of vibrational energy. Third, the acoustic phonon pedestal features two peaks next to the zero-phonon line (seen more clearly in Fig. \ref{Fig3}). Although the zero-frequency cutoff of the spectral density may result in a sharp feature on the Stokes-side ($\hbar\omega_t - \hbar\omega_\tau < 0$) of the zero-phonon line, we cannot explain the anti-Stokes ($\hbar\omega_t - \hbar\omega_\tau > 0$) peak (marked by red arrows in Fig. \ref{Fig3}) through solely an acoustic phonon continuum (see Supplemental Info).

We explain these results by proposing that discrete torsional modes related to the quantum dot geometry exist in conjunction with continuum longitudinal acoustic modes related to the crystal lattice. We were unable to find values for the low-temperature (glass phase) sound velocity for heptamethylnonane, though from its room temperature (liquid phase) speed $1.285\times 10^5$ cm/s \cite{LuningPrak2015} it is plausible that in solid form its longitudinal sound velocity may increase three- to four-fold and become comparable to the CQD longitudinal sound velocity (see Supplemental Info). In this case longitudinal vibrations may propagate into the surrounding glass matrix with minimal reflection at the dot boundary, giving rise to a continuum of longitudinal acoustic phonon modes. Simultaneously, transverse torsional modes involving no radial displacement into the surrounding matrix are supported by the dot geometry.

To investigate the validity of this interpretation, we simulate the acquired single-quantum spectra according to the following model. A discrete transverse torsional mode ``dresses" the ground and excited electronic states to generate ladders of states separated by the torsional mode energy \cite{May2011} (shown in Fig. \ref{Fig1}b). By calculating allowed torsional mode energies (indexed by $\ell$ and $n$) according to our material parameters, we find that the $(\ell,n) = (2,0)$ mode matches the features of our experimental spectra (see Supplemental Info). We note that the higher-frequency $(\ell,n) = (1,0)$ mode, which involves oscillations of a core and shell layer in opposite directions, should couple weakly to the type-I CdSe/CdZnS quantum dots studied here that confine both carriers in the CdSe core \cite{Lim2014}. The transition strengths between these states are determined by the Huang-Rhys parameter $S$ \cite{deJong2015}. The energies of these states are then modulated by a continuum bath of longitudinal acoustic phonons via elastic interactions. Assuming the bath is harmonic, that is the coupling is taken to be linear coupling to a continuous distribution of harmonic oscillators, we can characterize the system-bath interaction by a spectral density function $J(\omega)$ (shown in Fig. \ref{Fig1}c). The spectral density may be loosely interpreted as the frequency spectrum of the energy gap modulation by a coupled harmonic bath. For a spherical quantum dot with identical electron and hole localization radii, the spectral density of an acoustic phonon bath:
\begin{align}
    J_{a}(\omega) = A\omega^p e^{-\omega^2/\omega_c^2}
\end{align}
may be derived analytically \cite{Luker2017}, where $A$ characterizes the coupling strength, $\omega_c$ is a cutoff frequency that determines the width of acoustic phonon spectral features, and $p$ is an integer that depends on the coupling mechanism ($p = (1)3$ for (deformation potential) piezoelectric coupling) \cite{Calarco2003}. The experimental and simulated single-quantum spectra at temperatures 5, 10, and 16 Kelvin are shown in Fig. \ref{Fig2}. Corresponding simulations without torsional mode coupling, which fail to reproduce the anti-Stokes peak, may be found in the Supplemental Info. 

For the model described above, analytic expressions are not available to fit experimental lineshapes. We thus emphasize that the goal of our simulations is not to extract numerical values for specific quantities, but rather to elucidate the nature of the underlying microscopic dynamics. The torsional mode energy $E_{(2,0)}$ is calculated \cite{Takagahara1993}, while accounting for the core-shell structure of our dots (see Supplemental Info), to be 0.8 meV, and is the value used in our simulation. The torsional mode Huang-Rhys parameter used is $S = 0.6$. We were unable to obtain good agreement between experiment and simulation for deformation potential coupling ($p = 1$) with the continuum acoustic modes. Although excitons in free-standing few-nm radii CQDs are thought to couple to acoustic vibrations predominately via the deformation potential mechanism \cite{Takagahara1993}, our spectra suggest that the delocalized acoustic vibrations considered here couple through the long-range piezoelectric interaction ($p = 3$) used in our simulation. Coherences between states $\Ket{g}$ and $\Ket{e}$ are simulated with a dephasing rate $\hbar\gamma$ = 0.4 meV to match the zero-phonon linewidth while coherences involving the vibrationally-excited states $\Ket{\tilde{g}}$ and $\Ket{\tilde{e}}$ dephase more quickly at $\hbar\gamma$ = 1.6 meV. The remaining parameters used for the spectral density are $A = 0.47$ ps$^4$ and $\hbar\omega_c = 1.15$ meV. It should be noted that the simulated spectra in Fig. \ref{Fig2} are obtained after numerically including finite pulse bandwidth effects (see Supplemental Info) for a large inhomogeneously broadened ($\sigma = 50$ meV) response function by using the respective experimental laser spectrum at each temperature.

In Fig. \ref{Fig3}, we plot cross-diagonal slices centered at energy $|\hbar\omega_\tau| = |\hbar\omega_t|$ = 2060 meV. Comparison of the single-quantum spectrum slices with the simulated absorption and fluorescence lineshapes plotted inset contrasts the difference in strength (relative to the zero-phonon line) between linear and third-order vibrational lineshapes. The enhancement of vibrational lineshapes in the third-order response is due to additional terms involving the dephasing lineshape function which are absent in the linear response (see Supplemental Info). The nonlinear response of a system may therefore reveal vibrational couplings that are otherwise too weak to observe via linear spectroscopies.

In conclusion, we have observed vibrational lineshapes in the third-order nonlinear response of a core-shell CQD ensemble that indicate simultaneous existence of discrete and continuum acoustic vibrational modes. As a primary mechanism of energy loss and dephasing, understanding acoustic phonon coupling is crucial to the design and implementation of CQDs in optoelectronic devices. In particular, devices based on CQD-doped glasses \cite{Dong2015} or superlattices \cite{Kagan2016} should be designed with an engineered spectral density to minimize energy loss and other dissipative processes due to interactions with acoustic vibrations.

\begin{acknowledgement}

This work was supported by the Department of Energy grant number DE-SC0015782. D.B.A. acknowledges support by a fellowship from the Brazilian National Council for Scientific and Technological Development (CNPq). L.A.P. acknowledges support from FAPESP (Project numbers 2013/16911-2 and 2016/50011-7).

\end{acknowledgement}


\bibliography{bibliography}

\providecommand{\latin}[1]{#1}
\makeatletter
\providecommand{\doi}
  {\begingroup\let\do\@makeother\dospecials
  \catcode`\{=1 \catcode`\}=2 \doi@aux}
\providecommand{\doi@aux}[1]{\endgroup\texttt{#1}}
\makeatother
\providecommand*\mcitethebibliography{\thebibliography}
\csname @ifundefined\endcsname{endmcitethebibliography}
  {\let\endmcitethebibliography\endthebibliography}{}
\begin{mcitethebibliography}{53}
\providecommand*\natexlab[1]{#1}
\providecommand*\mciteSetBstSublistMode[1]{}
\providecommand*\mciteSetBstMaxWidthForm[2]{}
\providecommand*\mciteBstWouldAddEndPuncttrue
  {\def\EndOfBibitem{\unskip.}}
\providecommand*\mciteBstWouldAddEndPunctfalse
  {\let\EndOfBibitem\relax}
\providecommand*\mciteSetBstMidEndSepPunct[3]{}
\providecommand*\mciteSetBstSublistLabelBeginEnd[3]{}
\providecommand*\EndOfBibitem{}
\mciteSetBstSublistMode{f}
\mciteSetBstMaxWidthForm{subitem}{(\alph{mcitesubitemcount})}
\mciteSetBstSublistLabelBeginEnd
  {\mcitemaxwidthsubitemform\space}
  {\relax}
  {\relax}

\bibitem[Yuan \latin{et~al.}(2002)Yuan, Kardynal, Stevenson, Shields, Lobo,
  Cooper, Beattie, Ritchie, and Pepper]{Yuan2002}
Yuan,~Z.; Kardynal,~B.~E.; Stevenson,~R.~M.; Shields,~A.~J.; Lobo,~C.~J.;
  Cooper,~K.; Beattie,~N.~S.; Ritchie,~D.~A.; Pepper,~M. Electrically Driven
  Single-Photon Source. \emph{Science} \textbf{2002}, \emph{295}, 102\relax
\mciteBstWouldAddEndPuncttrue
\mciteSetBstMidEndSepPunct{\mcitedefaultmidpunct}
{\mcitedefaultendpunct}{\mcitedefaultseppunct}\relax
\EndOfBibitem
\bibitem[Senellart \latin{et~al.}(2017)Senellart, Solomon, and
  White]{Senellart2017}
Senellart,~P.; Solomon,~G.; White,~A. High-performance semiconductor
  quantum-dot single-photon sources. \emph{Nat. Nanotechnol.} \textbf{2017},
  \emph{12}, 1026, Review Article\relax
\mciteBstWouldAddEndPuncttrue
\mciteSetBstMidEndSepPunct{\mcitedefaultmidpunct}
{\mcitedefaultendpunct}{\mcitedefaultseppunct}\relax
\EndOfBibitem
\bibitem[Loss and DiVincenzo(1998)Loss, and DiVincenzo]{Loss1998}
Loss,~D.; DiVincenzo,~D.~P. Quantum computation with quantum dots. \emph{Phys.
  Rev. A} \textbf{1998}, \emph{57}, 120--126\relax
\mciteBstWouldAddEndPuncttrue
\mciteSetBstMidEndSepPunct{\mcitedefaultmidpunct}
{\mcitedefaultendpunct}{\mcitedefaultseppunct}\relax
\EndOfBibitem
\bibitem[Kloeffel and Loss(2013)Kloeffel, and Loss]{Kloeffel2013}
Kloeffel,~C.; Loss,~D. Prospects for Spin-Based Quantum Computing in Quantum
  Dots. \emph{Annu. Rev. Condens. Matter Phys.} \textbf{2013}, \emph{4},
  51--81\relax
\mciteBstWouldAddEndPuncttrue
\mciteSetBstMidEndSepPunct{\mcitedefaultmidpunct}
{\mcitedefaultendpunct}{\mcitedefaultseppunct}\relax
\EndOfBibitem
\bibitem[Gazzano and Solomon(2016)Gazzano, and Solomon]{Gazzano2016}
Gazzano,~O.; Solomon,~G.~S. Toward optical quantum information processing with
  quantum dots coupled to microstructures [Invited]. \emph{J. Opt. Soc. Am. B}
  \textbf{2016}, \emph{33}, C160--C175\relax
\mciteBstWouldAddEndPuncttrue
\mciteSetBstMidEndSepPunct{\mcitedefaultmidpunct}
{\mcitedefaultendpunct}{\mcitedefaultseppunct}\relax
\EndOfBibitem
\bibitem[Carey \latin{et~al.}(2015)Carey, Abdelhady, Ning, Thon, Bakr, and
  Sargent]{Carey2015}
Carey,~G.~H.; Abdelhady,~A.~L.; Ning,~Z.; Thon,~S.~M.; Bakr,~O.~M.;
  Sargent,~E.~H. Colloidal Quantum Dot Solar Cells. \emph{Chem. Rev.}
  \textbf{2015}, \emph{115}, 12732--12763\relax
\mciteBstWouldAddEndPuncttrue
\mciteSetBstMidEndSepPunct{\mcitedefaultmidpunct}
{\mcitedefaultendpunct}{\mcitedefaultseppunct}\relax
\EndOfBibitem
\bibitem[Zhao and Rosei(2017)Zhao, and Rosei]{Zhao2017}
Zhao,~H.; Rosei,~F. Colloidal Quantum Dots for Solar Technologies. \emph{Chem}
  \textbf{2017}, \emph{3}, 229--258\relax
\mciteBstWouldAddEndPuncttrue
\mciteSetBstMidEndSepPunct{\mcitedefaultmidpunct}
{\mcitedefaultendpunct}{\mcitedefaultseppunct}\relax
\EndOfBibitem
\bibitem[Quilter \latin{et~al.}(2015)Quilter, Brash, Liu, Gl{\"a}ssl, Barth,
  Axt, Ramsay, Skolnick, and Fox]{Quilter2015}
Quilter,~J.~H.; Brash,~A.~J.; Liu,~F.; Gl{\"a}ssl,~M.; Barth,~A.~M.;
  Axt,~V.~M.; Ramsay,~A.~J.; Skolnick,~M.~S.; Fox,~A.~M. Phonon-Assisted
  Population Inversion of a Single InGaAs/GaAs Quantum Dot by Pulsed Laser
  Excitation. \emph{Phys. Rev. Lett.} \textbf{2015}, \emph{114}, 137401\relax
\mciteBstWouldAddEndPuncttrue
\mciteSetBstMidEndSepPunct{\mcitedefaultmidpunct}
{\mcitedefaultendpunct}{\mcitedefaultseppunct}\relax
\EndOfBibitem
\bibitem[Kaldewey \latin{et~al.}(2017)Kaldewey, L{\"u}ker, Kuhlmann, Valentin,
  Ludwig, Wieck, Reiter, Kuhn, and Warburton]{Kaldewey2017}
Kaldewey,~T.; L{\"u}ker,~S.; Kuhlmann,~A.~V.; Valentin,~S.~R.; Ludwig,~A.;
  Wieck,~A.~D.; Reiter,~D.~E.; Kuhn,~T.; Warburton,~R.~J. Coherent and robust
  high-fidelity generation of a biexciton in a quantum dot by rapid adiabatic
  passage. \emph{Phys. Rev. B} \textbf{2017}, \emph{95}, 161302\relax
\mciteBstWouldAddEndPuncttrue
\mciteSetBstMidEndSepPunct{\mcitedefaultmidpunct}
{\mcitedefaultendpunct}{\mcitedefaultseppunct}\relax
\EndOfBibitem
\bibitem[Reindl \latin{et~al.}(2017)Reindl, J{\"o}ns, Huber, Schimpf, Huo,
  Zwiller, Rastelli, and Trotta]{Reindl2017}
Reindl,~M.; J{\"o}ns,~K.~D.; Huber,~D.; Schimpf,~C.; Huo,~Y.; Zwiller,~V.;
  Rastelli,~A.; Trotta,~R. Phonon-Assisted Two-Photon Interference from Remote
  Quantum Emitters. \emph{Nano Lett.} \textbf{2017}, \emph{17},
  4090--4095\relax
\mciteBstWouldAddEndPuncttrue
\mciteSetBstMidEndSepPunct{\mcitedefaultmidpunct}
{\mcitedefaultendpunct}{\mcitedefaultseppunct}\relax
\EndOfBibitem
\bibitem[Poem \latin{et~al.}(2010)Poem, Kodriano, Tradonsky, Lindner, Gerardot,
  Petroff, and Gershoni]{Poem2010}
Poem,~E.; Kodriano,~Y.; Tradonsky,~C.; Lindner,~N.~H.; Gerardot,~B.~D.;
  Petroff,~P.~M.; Gershoni,~D. Accessing the dark exciton with light.
  \emph{Nat. Phys.} \textbf{2010}, \emph{6}, Article\relax
\mciteBstWouldAddEndPuncttrue
\mciteSetBstMidEndSepPunct{\mcitedefaultmidpunct}
{\mcitedefaultendpunct}{\mcitedefaultseppunct}\relax
\EndOfBibitem
\bibitem[Masia \latin{et~al.}(2012)Masia, Accanto, Langbein, and
  Borri]{Masia2012}
Masia,~F.; Accanto,~N.; Langbein,~W.; Borri,~P. Spin-Flip Limited Exciton
  Dephasing in CdSe/ZnS Colloidal Quantum Dots. \emph{Phys. Rev. Lett.}
  \textbf{2012}, \emph{108}, 087401\relax
\mciteBstWouldAddEndPuncttrue
\mciteSetBstMidEndSepPunct{\mcitedefaultmidpunct}
{\mcitedefaultendpunct}{\mcitedefaultseppunct}\relax
\EndOfBibitem
\bibitem[L{\"u}ker \latin{et~al.}(2017)L{\"u}ker, Kuhn, and
  Reiter]{Luker2017-2}
L{\"u}ker,~S.; Kuhn,~T.; Reiter,~D.~E. Phonon-assisted dark exciton preparation
  in a quantum dot. \emph{Phys. Rev. B} \textbf{2017}, \emph{95}, 195305\relax
\mciteBstWouldAddEndPuncttrue
\mciteSetBstMidEndSepPunct{\mcitedefaultmidpunct}
{\mcitedefaultendpunct}{\mcitedefaultseppunct}\relax
\EndOfBibitem
\bibitem[Muljarov and Zimmermann(2004)Muljarov, and Zimmermann]{Muljarov2004}
Muljarov,~E.~A.; Zimmermann,~R. Dephasing in Quantum Dots: Quadratic Coupling
  to Acoustic Phonons. \emph{Phys. Rev. Lett.} \textbf{2004}, \emph{93},
  237401\relax
\mciteBstWouldAddEndPuncttrue
\mciteSetBstMidEndSepPunct{\mcitedefaultmidpunct}
{\mcitedefaultendpunct}{\mcitedefaultseppunct}\relax
\EndOfBibitem
\bibitem[Accanto \latin{et~al.}(2012)Accanto, Masia, Moreels, Hens, Langbein,
  and Borri]{Accanto2012}
Accanto,~N.; Masia,~F.; Moreels,~I.; Hens,~Z.; Langbein,~W.; Borri,~P.
  Engineering the Spin-Flip Limited Exciton Dephasing in Colloidal CdSe/CdS
  Quantum Dots. \emph{ACS Nano} \textbf{2012}, \emph{6}, 5227--5233\relax
\mciteBstWouldAddEndPuncttrue
\mciteSetBstMidEndSepPunct{\mcitedefaultmidpunct}
{\mcitedefaultendpunct}{\mcitedefaultseppunct}\relax
\EndOfBibitem
\bibitem[Jakubczyk \latin{et~al.}(2016)Jakubczyk, Delmonte, Fischbach, Wigger,
  Reiter, Mermillod, Schnauber, Kaganskiy, Schulze, Strittmatter, Rodt,
  Langbein, Kuhn, Reitzenstein, and Kasprzak]{Jakubczyk2016}
Jakubczyk,~T.; Delmonte,~V.; Fischbach,~S.; Wigger,~D.; Reiter,~D.~E.;
  Mermillod,~Q.; Schnauber,~P.; Kaganskiy,~A.; Schulze,~J.-H.; Strittmatter,~A.
  \latin{et~al.}  Impact of Phonons on Dephasing of Individual Excitons in
  Deterministic Quantum Dot Microlenses. \emph{ACS Photonics} \textbf{2016},
  \emph{3}, 2461--2466\relax
\mciteBstWouldAddEndPuncttrue
\mciteSetBstMidEndSepPunct{\mcitedefaultmidpunct}
{\mcitedefaultendpunct}{\mcitedefaultseppunct}\relax
\EndOfBibitem
\bibitem[Takagahara(1993)]{Takagahara1993}
Takagahara,~T. Electron-phonon interactions and excitonic dephasing in
  semiconductor nanocrystals. \emph{Phys. Rev. Lett.} \textbf{1993}, \emph{71},
  3577--3580\relax
\mciteBstWouldAddEndPuncttrue
\mciteSetBstMidEndSepPunct{\mcitedefaultmidpunct}
{\mcitedefaultendpunct}{\mcitedefaultseppunct}\relax
\EndOfBibitem
\bibitem[Kuok \latin{et~al.}(2003)Kuok, Lim, Ng, Liu, and Wang]{Kuok2003}
Kuok,~M.~H.; Lim,~H.~S.; Ng,~S.~C.; Liu,~N.~N.; Wang,~Z.~K. Brillouin Study of
  the Quantization of Acoustic Modes in Nanospheres. \emph{Phys. Rev. Lett.}
  \textbf{2003}, \emph{90}, 255502\relax
\mciteBstWouldAddEndPuncttrue
\mciteSetBstMidEndSepPunct{\mcitedefaultmidpunct}
{\mcitedefaultendpunct}{\mcitedefaultseppunct}\relax
\EndOfBibitem
\bibitem[Chilla \latin{et~al.}(2008)Chilla, Kipp, Menke, Heitmann, Nikolic,
  Fr{\"o}msdorf, Kornowski, F{\"o}rster, and Weller]{Chilla2008}
Chilla,~G.; Kipp,~T.; Menke,~T.; Heitmann,~D.; Nikolic,~M.; Fr{\"o}msdorf,~A.;
  Kornowski,~A.; F{\"o}rster,~S.; Weller,~H. Direct Observation of Confined
  Acoustic Phonons in the Photoluminescence Spectra of a Single CdSe-CdS-ZnS
  Core-Shell-Shell Nanocrystal. \emph{Phys. Rev. Lett.} \textbf{2008},
  \emph{100}, 057403\relax
\mciteBstWouldAddEndPuncttrue
\mciteSetBstMidEndSepPunct{\mcitedefaultmidpunct}
{\mcitedefaultendpunct}{\mcitedefaultseppunct}\relax
\EndOfBibitem
\bibitem[Oron \latin{et~al.}(2009)Oron, Aharoni, de~Mello~Donega, van Rijssel,
  Meijerink, and Banin]{Oron2009}
Oron,~D.; Aharoni,~A.; de~Mello~Donega,~C.; van Rijssel,~J.; Meijerink,~A.;
  Banin,~U. Universal Role of Discrete Acoustic Phonons in the Low-Temperature
  Optical Emission of Colloidal Quantum Dots. \emph{Phys. Rev. Lett.}
  \textbf{2009}, \emph{102}, 177402\relax
\mciteBstWouldAddEndPuncttrue
\mciteSetBstMidEndSepPunct{\mcitedefaultmidpunct}
{\mcitedefaultendpunct}{\mcitedefaultseppunct}\relax
\EndOfBibitem
\bibitem[Huxter and Scholes(2010)Huxter, and Scholes]{Huxter2010}
Huxter,~V.~M.; Scholes,~G.~D. Acoustic phonon strain induced mixing of the fine
  structure levels in colloidal CdSe quantum dots observed by a polarization
  grating technique. \emph{J. Chem. Phys.} \textbf{2010}, \emph{132},
  104506\relax
\mciteBstWouldAddEndPuncttrue
\mciteSetBstMidEndSepPunct{\mcitedefaultmidpunct}
{\mcitedefaultendpunct}{\mcitedefaultseppunct}\relax
\EndOfBibitem
\bibitem[Werschler \latin{et~al.}(2016)Werschler, Hinz, Froning, Gumbsheimer,
  Haase, Negele, de~Roo, Mecking, Leitenstorfer, and Seletskiy]{Werschler2016}
Werschler,~F.; Hinz,~C.; Froning,~F.; Gumbsheimer,~P.; Haase,~J.; Negele,~C.;
  de~Roo,~T.; Mecking,~S.; Leitenstorfer,~A.; Seletskiy,~D.~V. Coupling of
  Excitons and Discrete Acoustic Phonons in Vibrationally Isolated Quantum
  Emitters. \emph{Nano Lett.} \textbf{2016}, \emph{16}, 5861--5865\relax
\mciteBstWouldAddEndPuncttrue
\mciteSetBstMidEndSepPunct{\mcitedefaultmidpunct}
{\mcitedefaultendpunct}{\mcitedefaultseppunct}\relax
\EndOfBibitem
\bibitem[Empedocles and Bawendi(1999)Empedocles, and Bawendi]{Empedocles1999}
Empedocles,~S.~A.; Bawendi,~M.~G. Influence of Spectral Diffusion on the Line
  Shapes of Single CdSe Nanocrystallite Quantum Dots. \emph{J. Phys. Chem. B}
  \textbf{1999}, \emph{103}, 1826--1830\relax
\mciteBstWouldAddEndPuncttrue
\mciteSetBstMidEndSepPunct{\mcitedefaultmidpunct}
{\mcitedefaultendpunct}{\mcitedefaultseppunct}\relax
\EndOfBibitem
\bibitem[Fern{\'e}e \latin{et~al.}(2008)Fern{\'e}e, Littleton, Cooper,
  Rubinsztein-Dunlop, G{\'o}mez, and Mulvaney]{Fernee2008}
Fern{\'e}e,~M.~J.; Littleton,~B.~N.; Cooper,~S.; Rubinsztein-Dunlop,~H.;
  G{\'o}mez,~D.~E.; Mulvaney,~P. Acoustic Phonon Contributions to the Emission
  Spectrum of Single CdSe Nanocrystals. \emph{J. Phys. Chem. C} \textbf{2008},
  \emph{112}, 1878--1884\relax
\mciteBstWouldAddEndPuncttrue
\mciteSetBstMidEndSepPunct{\mcitedefaultmidpunct}
{\mcitedefaultendpunct}{\mcitedefaultseppunct}\relax
\EndOfBibitem
\bibitem[Cundiff and Mukamel(2013)Cundiff, and Mukamel]{Cundiff2013}
Cundiff,~S.~T.; Mukamel,~S. Optical multidimensional coherent spectroscopy.
  \emph{Phys. Today} \textbf{2013}, \emph{66}, 44--49\relax
\mciteBstWouldAddEndPuncttrue
\mciteSetBstMidEndSepPunct{\mcitedefaultmidpunct}
{\mcitedefaultendpunct}{\mcitedefaultseppunct}\relax
\EndOfBibitem
\bibitem[Ferrio and Steel(1998)Ferrio, and Steel]{Ferrio1998}
Ferrio,~K.~B.; Steel,~D.~G. Raman Quantum Beats of Interacting Excitons.
  \emph{Phys. Rev. Lett.} \textbf{1998}, \emph{80}, 786--789\relax
\mciteBstWouldAddEndPuncttrue
\mciteSetBstMidEndSepPunct{\mcitedefaultmidpunct}
{\mcitedefaultendpunct}{\mcitedefaultseppunct}\relax
\EndOfBibitem
\bibitem[Kasprzak \latin{et~al.}(2010)Kasprzak, Patton, Savona, and
  Langbein]{Kasprzak2010}
Kasprzak,~J.; Patton,~B.; Savona,~V.; Langbein,~W. Coherent coupling between
  distant excitons revealed by two-dimensional nonlinear hyperspectral imaging.
  \emph{Nat. Photonics} \textbf{2010}, \emph{5}, Article\relax
\mciteBstWouldAddEndPuncttrue
\mciteSetBstMidEndSepPunct{\mcitedefaultmidpunct}
{\mcitedefaultendpunct}{\mcitedefaultseppunct}\relax
\EndOfBibitem
\bibitem[Martin and Cundiff(2018)Martin, and Cundiff]{Martin2018}
Martin,~E.~W.; Cundiff,~S.~T. Inducing coherent quantum dot interactions.
  \emph{Phys. Rev. B} \textbf{2018}, \emph{97}, 081301\relax
\mciteBstWouldAddEndPuncttrue
\mciteSetBstMidEndSepPunct{\mcitedefaultmidpunct}
{\mcitedefaultendpunct}{\mcitedefaultseppunct}\relax
\EndOfBibitem
\bibitem[Langbein \latin{et~al.}(2013)Langbein, Portolan, Rastelli, Wang,
  Plumhof, Schmidt, and W.]{Langbein2013}
Langbein,~J.~K.; Portolan,~S.; Rastelli,~A.; Wang,~L.; Plumhof,~J.~D.;
  Schmidt,~O.~G.; W., Vectorial nonlinear coherent response of a strongly
  confined exciton-biexciton system. \emph{New J. Phys.} \textbf{2013},
  \emph{15}, 055006\relax
\mciteBstWouldAddEndPuncttrue
\mciteSetBstMidEndSepPunct{\mcitedefaultmidpunct}
{\mcitedefaultendpunct}{\mcitedefaultseppunct}\relax
\EndOfBibitem
\bibitem[Suzuki \latin{et~al.}(2016)Suzuki, Singh, Bayer, Ludwig, Wieck, and
  Cundiff]{Suzuki2016}
Suzuki,~T.; Singh,~R.; Bayer,~M.; Ludwig,~A.; Wieck,~A.~D.; Cundiff,~S.~T.
  Coherent Control of the Exciton-Biexciton System in an InAs Self-Assembled
  Quantum Dot Ensemble. \emph{Phys. Rev. Lett.} \textbf{2016}, \emph{117},
  157402\relax
\mciteBstWouldAddEndPuncttrue
\mciteSetBstMidEndSepPunct{\mcitedefaultmidpunct}
{\mcitedefaultendpunct}{\mcitedefaultseppunct}\relax
\EndOfBibitem
\bibitem[Harel \latin{et~al.}(2012)Harel, Rupich, Schaller, Talapin, and
  Engel]{Harel2012}
Harel,~E.; Rupich,~S.~M.; Schaller,~R.~D.; Talapin,~D.~V.; Engel,~G.~S.
  Measurement of electronic splitting in PbS quantum dots by two-dimensional
  nonlinear spectroscopy. \emph{Physical Review B} \textbf{2012}, \emph{86},
  075412\relax
\mciteBstWouldAddEndPuncttrue
\mciteSetBstMidEndSepPunct{\mcitedefaultmidpunct}
{\mcitedefaultendpunct}{\mcitedefaultseppunct}\relax
\EndOfBibitem
\bibitem[Karki \latin{et~al.}(2014)Karki, Widom, Seibt, Moody, Lonergan,
  Pullerits, and Marcus]{Karki2014}
Karki,~K.~J.; Widom,~J.~R.; Seibt,~J.; Moody,~I.; Lonergan,~M.~C.;
  Pullerits,~T.; Marcus,~A.~H. Coherent two-dimensional photocurrent
  spectroscopy in a PbS quantum dot photocell. \emph{Nat. Commun.}
  \textbf{2014}, \emph{5}, 5869, Article\relax
\mciteBstWouldAddEndPuncttrue
\mciteSetBstMidEndSepPunct{\mcitedefaultmidpunct}
{\mcitedefaultendpunct}{\mcitedefaultseppunct}\relax
\EndOfBibitem
\bibitem[Gellen \latin{et~al.}(2017)Gellen, Lem, and Turner]{Gellen2017}
Gellen,~T.~A.; Lem,~J.; Turner,~D.~B. Probing Homogeneous Line Broadening in
  CdSe Nanocrystals Using Multidimensional Electronic Spectroscopy. \emph{Nano
  Lett.} \textbf{2017}, \emph{17}, 2809--2815\relax
\mciteBstWouldAddEndPuncttrue
\mciteSetBstMidEndSepPunct{\mcitedefaultmidpunct}
{\mcitedefaultendpunct}{\mcitedefaultseppunct}\relax
\EndOfBibitem
\bibitem[Park \latin{et~al.}(2017)Park, Baranov, Ryu, Cho, Halder, Seifert,
  Vajda, and Jonas]{Park2017}
Park,~S.~D.; Baranov,~D.; Ryu,~J.; Cho,~B.; Halder,~A.; Seifert,~S.; Vajda,~S.;
  Jonas,~D.~M. Bandgap Inhomogeneity of a PbSe Quantum Dot Ensemble from
  Two-Dimensional Spectroscopy and Comparison to Size Inhomogeneity from
  Electron Microscopy. \emph{Nano Lett.} \textbf{2017}, \emph{17},
  762--771\relax
\mciteBstWouldAddEndPuncttrue
\mciteSetBstMidEndSepPunct{\mcitedefaultmidpunct}
{\mcitedefaultendpunct}{\mcitedefaultseppunct}\relax
\EndOfBibitem
\bibitem[Seiler \latin{et~al.}(2018)Seiler, Palato, Sonnichsen, Baker, and
  Kambhampati]{Seiler2018}
Seiler,~H.; Palato,~S.; Sonnichsen,~C.; Baker,~H.; Kambhampati,~P. Seeing
  Multiexcitons through Sample Inhomogeneity: Band-Edge Biexciton Structure in
  CdSe Nanocrystals Revealed by Two-Dimensional Electronic Spectroscopy.
  \emph{Nano Lett.} \textbf{2018}, \emph{18}, 2999--3006\relax
\mciteBstWouldAddEndPuncttrue
\mciteSetBstMidEndSepPunct{\mcitedefaultmidpunct}
{\mcitedefaultendpunct}{\mcitedefaultseppunct}\relax
\EndOfBibitem
\bibitem[{Liu} \latin{et~al.}(2019){Liu}, {Almeida}, {Bae}, {Padilha}, and
  {Cundiff}]{Liu2018}
{Liu},~A.; {Almeida},~D.~B.; {Bae},~W.~K.; {Padilha},~L.~A.; {Cundiff},~S.~T.
  {Vibrational Coupling Modifies Spectral Diffusion in Core-Shell Colloidal
  Quantum Dots}. \emph{ArXiv:1806.06112 (accepted for publication in Physical
  Review Letters)} \textbf{2019}, \relax
\mciteBstWouldAddEndPunctfalse
\mciteSetBstMidEndSepPunct{\mcitedefaultmidpunct}
{}{\mcitedefaultseppunct}\relax
\EndOfBibitem
\bibitem[Sirenko \latin{et~al.}(1998)Sirenko, Belitsky, Ruf, Cardona, Ekimov,
  and Trallero-Giner]{Sirenko1998}
Sirenko,~A.~A.; Belitsky,~V.~I.; Ruf,~T.; Cardona,~M.; Ekimov,~A.~I.;
  Trallero-Giner,~C. Spin-flip and acoustic-phonon Raman scattering in CdS
  nanocrystals. \emph{Phys. Rev. B} \textbf{1998}, \emph{58}, 2077--2087\relax
\mciteBstWouldAddEndPuncttrue
\mciteSetBstMidEndSepPunct{\mcitedefaultmidpunct}
{\mcitedefaultendpunct}{\mcitedefaultseppunct}\relax
\EndOfBibitem
\bibitem[Murray and Saviot(2004)Murray, and Saviot]{Murray2004}
Murray,~D.~B.; Saviot,~L. Phonons in an inhomogeneous continuum: Vibrations of
  an embedded nanoparticle. \emph{Phys. Rev. B} \textbf{2004}, \emph{69},
  094305\relax
\mciteBstWouldAddEndPuncttrue
\mciteSetBstMidEndSepPunct{\mcitedefaultmidpunct}
{\mcitedefaultendpunct}{\mcitedefaultseppunct}\relax
\EndOfBibitem
\bibitem[Lim \latin{et~al.}(2014)Lim, Jeong, Park, Kim, Pietryga, Park, Klimov,
  Lee, Lee, and Bae]{Lim2014}
Lim,~J.; Jeong,~B.~G.; Park,~M.; Kim,~J.~K.; Pietryga,~J.~M.; Park,~Y.-S.;
  Klimov,~V.~I.; Lee,~C.; Lee,~D.~C.; Bae,~W.~K. Influence of Shell Thickness
  on the Performance of Light-Emitting Devices Based on CdSe/Zn1-XCdXS
  Core/Shell Heterostructured Quantum Dots. \emph{Adv. Mater.} \textbf{2014},
  \emph{26}, 8034--8040\relax
\mciteBstWouldAddEndPuncttrue
\mciteSetBstMidEndSepPunct{\mcitedefaultmidpunct}
{\mcitedefaultendpunct}{\mcitedefaultseppunct}\relax
\EndOfBibitem
\bibitem[Hamm and Zanni(2011)Hamm, and Zanni]{Hamm2011}
Hamm,~P.; Zanni,~M. \emph{Concepts and Methods of 2D Infrared Spectroscopy},
  1st ed.; Cambridge University Press, 2011\relax
\mciteBstWouldAddEndPuncttrue
\mciteSetBstMidEndSepPunct{\mcitedefaultmidpunct}
{\mcitedefaultendpunct}{\mcitedefaultseppunct}\relax
\EndOfBibitem
\bibitem[Bristow \latin{et~al.}(2009)Bristow, Karaiskaj, Dai, Zhang, Carlsson,
  Hagen, Jimenez, and Cundiff]{Bristow2009}
Bristow,~A.~D.; Karaiskaj,~D.; Dai,~X.; Zhang,~T.; Carlsson,~C.; Hagen,~K.~R.;
  Jimenez,~R.; Cundiff,~S.~T. A versatile ultrastable platform for optical
  multidimensional Fourier-transform spectroscopy. \emph{Rev. Sci. Instrum.}
  \textbf{2009}, \emph{80}, 073108\relax
\mciteBstWouldAddEndPuncttrue
\mciteSetBstMidEndSepPunct{\mcitedefaultmidpunct}
{\mcitedefaultendpunct}{\mcitedefaultseppunct}\relax
\EndOfBibitem
\bibitem[Palinginis and Wang(2001)Palinginis, and Wang]{Palinginis2001}
Palinginis,~P.; Wang,~H. High-resolution spectral hole burning in CdSe/ZnS
  core/shell nanocrystals. \emph{Appl. Phys. Lett.} \textbf{2001}, \emph{78},
  1541--1543\relax
\mciteBstWouldAddEndPuncttrue
\mciteSetBstMidEndSepPunct{\mcitedefaultmidpunct}
{\mcitedefaultendpunct}{\mcitedefaultseppunct}\relax
\EndOfBibitem
\bibitem[Palinginis \latin{et~al.}(2003)Palinginis, Tavenner, Lonergan, and
  Wang]{Palinginis2003}
Palinginis,~P.; Tavenner,~S.; Lonergan,~M.; Wang,~H. Spectral hole burning and
  zero phonon linewidth in semiconductor nanocrystals. \emph{Phys. Rev. B}
  \textbf{2003}, \emph{67}, 201307\relax
\mciteBstWouldAddEndPuncttrue
\mciteSetBstMidEndSepPunct{\mcitedefaultmidpunct}
{\mcitedefaultendpunct}{\mcitedefaultseppunct}\relax
\EndOfBibitem
\bibitem[Krummheuer \latin{et~al.}(2002)Krummheuer, Axt, and
  Kuhn]{Krummheuer2002}
Krummheuer,~B.; Axt,~V.~M.; Kuhn,~T. Theory of pure dephasing and the resulting
  absorption line shape in semiconductor quantum dots. \emph{Phys. Rev. B}
  \textbf{2002}, \emph{65}, 195313\relax
\mciteBstWouldAddEndPuncttrue
\mciteSetBstMidEndSepPunct{\mcitedefaultmidpunct}
{\mcitedefaultendpunct}{\mcitedefaultseppunct}\relax
\EndOfBibitem
\bibitem[Besombes \latin{et~al.}(2001)Besombes, Kheng, Marsal, and
  Mariette]{Besombes2001}
Besombes,~L.; Kheng,~K.; Marsal,~L.; Mariette,~H. Acoustic phonon broadening
  mechanism in single quantum dot emission. \emph{Phys. Rev. B} \textbf{2001},
  \emph{63}, 155307\relax
\mciteBstWouldAddEndPuncttrue
\mciteSetBstMidEndSepPunct{\mcitedefaultmidpunct}
{\mcitedefaultendpunct}{\mcitedefaultseppunct}\relax
\EndOfBibitem
\bibitem[Luning~Prak \latin{et~al.}(2015)Luning~Prak, Jones, Cowart, and
  Trulove]{LuningPrak2015}
Luning~Prak,~D.~J.; Jones,~M.~H.; Cowart,~J.~S.; Trulove,~P.~C. Density,
  Viscosity, Speed of Sound, Bulk Modulus, Surface Tension, and Flash Point of
  Binary Mixtures of 2,2,4,6,6-Pentamethylheptane and
  2,2,4,4,6,8,8-Heptamethylnonane at (293.15 to 373.15) K and 0.1 MPa and
  Comparisons with Alcohol-to-Jet Fuel. \emph{J. Chem. Eng. Data}
  \textbf{2015}, \emph{60}, 1157--1165\relax
\mciteBstWouldAddEndPuncttrue
\mciteSetBstMidEndSepPunct{\mcitedefaultmidpunct}
{\mcitedefaultendpunct}{\mcitedefaultseppunct}\relax
\EndOfBibitem
\bibitem[May and K{\"u}hn(2011)May, and K{\"u}hn]{May2011}
May,~V.; K{\"u}hn,~O. \emph{Charge and Energy Transfer Dynamics in Molecular
  Systems}, 3rd ed.; Wiley-VCH, 2011\relax
\mciteBstWouldAddEndPuncttrue
\mciteSetBstMidEndSepPunct{\mcitedefaultmidpunct}
{\mcitedefaultendpunct}{\mcitedefaultseppunct}\relax
\EndOfBibitem
\bibitem[de~Jong \latin{et~al.}(2015)de~Jong, Seijo, Meijerink, and
  Rabouw]{deJong2015}
de~Jong,~M.; Seijo,~L.; Meijerink,~A.; Rabouw,~F.~T. Resolving the ambiguity in
  the relation between Stokes shift and Huang-Rhys parameter. \emph{Phys. Chem.
  Chem. Phys.} \textbf{2015}, \emph{17}, 16959--16969\relax
\mciteBstWouldAddEndPuncttrue
\mciteSetBstMidEndSepPunct{\mcitedefaultmidpunct}
{\mcitedefaultendpunct}{\mcitedefaultseppunct}\relax
\EndOfBibitem
\bibitem[L{\"u}ker \latin{et~al.}(2017)L{\"u}ker, Kuhn, and Reiter]{Luker2017}
L{\"u}ker,~S.; Kuhn,~T.; Reiter,~D.~E. Phonon impact on optical control schemes
  of quantum dots: Role of quantum dot geometry and symmetry. \emph{Phys. Rev.
  B} \textbf{2017}, \emph{96}, 245306\relax
\mciteBstWouldAddEndPuncttrue
\mciteSetBstMidEndSepPunct{\mcitedefaultmidpunct}
{\mcitedefaultendpunct}{\mcitedefaultseppunct}\relax
\EndOfBibitem
\bibitem[Calarco \latin{et~al.}(2003)Calarco, Datta, Fedichev, Pazy, and
  Zoller]{Calarco2003}
Calarco,~T.; Datta,~A.; Fedichev,~P.; Pazy,~E.; Zoller,~P. Spin-based
  all-optical quantum computation with quantum dots: Understanding and
  suppressing decoherence. \emph{Phys. Rev. A} \textbf{2003}, \emph{68},
  012310\relax
\mciteBstWouldAddEndPuncttrue
\mciteSetBstMidEndSepPunct{\mcitedefaultmidpunct}
{\mcitedefaultendpunct}{\mcitedefaultseppunct}\relax
\EndOfBibitem
\bibitem[Dong \latin{et~al.}(2015)Dong, Wang, Chen, Pan, and Qiu]{Dong2015}
Dong,~G.; Wang,~H.; Chen,~G.; Pan,~Q.; Qiu,~J. Quantum Dot-Doped Glasses and
  Fibers: Fabrication and Optical Properties. \emph{Front. Mater.}
  \textbf{2015}, \emph{2}, 13\relax
\mciteBstWouldAddEndPuncttrue
\mciteSetBstMidEndSepPunct{\mcitedefaultmidpunct}
{\mcitedefaultendpunct}{\mcitedefaultseppunct}\relax
\EndOfBibitem
\bibitem[Kagan \latin{et~al.}(2016)Kagan, Lifshitz, Sargent, and
  Talapin]{Kagan2016}
Kagan,~C.~R.; Lifshitz,~E.; Sargent,~E.~H.; Talapin,~D.~V. Building devices
  from colloidal quantum dots. \emph{Science} \textbf{2016}, \emph{353}\relax
\mciteBstWouldAddEndPuncttrue
\mciteSetBstMidEndSepPunct{\mcitedefaultmidpunct}
{\mcitedefaultendpunct}{\mcitedefaultseppunct}\relax
\EndOfBibitem
\end{mcitethebibliography}

\end{document}